\documentstyle[12pt]{article}
\begin{document}
\newcommand{\ev}[1]{\langle #1 \rangle}
\def\beq{\begin{equation}}
\def\eeq{\end{equation}}
\def\bnu{\bar{\nu}}
\def\footnoterule{\kern-3pt\hrule\kern3pt}
\newfont\figfont{cmr7 scaled 1200}


\renewcommand{\thefootnote}{\fnsymbol{footnote}}

\begin{center}
EUROPEAN ORGANIZATION FOR NUCLEAR RESEARCH
\end{center}
\rightline{CERN-PPE/94-07}
\rightline{MIT-CTP \#2270}
\rightline{hep-ph/9401299}
\rightline{January 1994}

\begin{center}
{\LARGE Optimized Variables \\
\vspace{5pt}
for the Study of $\Lambda_b$ Polarization%
\footnote{This work is supported
in part by funds provided by the U.S.
Department of Energy (DOE) under contract \#DE-AC02-76ER03069 and in
part by the Texas National Research Laboratory Commission
under grant \#RGFY92C6.\hfil\break
\vskip 0.05cm
\noindent $^{\dagger}$National
Science Foundation Young Investigator Award.\hfill\break
Alfred P.~Sloan
Foundation Research Fellowship.\hfil\break
Department of Energy Outstanding Junior
Investigator Award.\hfill\break}} \\
\vspace{20pt}
{\large G.~Bonvicini \\
CERN, {\it CH-1211 Geneva, Switzerland} \\
\vspace{10pt}
L.~Randall*$^\dagger$ \\
Mass. Inst. of Technology, {\it  Cambridge MA 02139, USA}} \\
\end{center}

\setcounter{page}{0}
\thispagestyle{empty}
\vspace{20pt}


\begin{abstract}
The value of the $b$-baryon polarization can be extracted from inclusive data
at LEP with better than 10\% precision based on current statistics. We present
a new variable by which to measure the polarization, which is the ratio of the
average electron energy to the average neutrino energy. This variable is both
sensitive to polarization and insensitive to fragmentation uncertainties.
\end{abstract}

\vspace{\fill}
\begin{center}
{\it Submitted to:} Physical Review Letters
\end{center}

\newpage
\renewcommand{\thefootnote}{\arabic{footnote}}
\setcounter{footnote}{0}

The problem of polarization transfer from the (heavy) quark produced in
$Z$ decay to the experimentally observed hadron has attracted
considerable interest in the last two
years \cite{close,korner,schuler,peskin,wise,alta}.
The $b$-quark coupling to the $Z$ is -94\% polarized according to the
Standard Model. At LEP $b$-quarks are
produced copiously in $Z$ decays
with an energy of $\sim 45$ GeV. \par
Because
the $\Lambda_b$ baryon, which
accounts for roughly 10\% of all
$b$-hadrons, retains
the initial $b$-quark spin if
produced directly,
one  expects a highly polarized $\Lambda_b$ sample.
Because higher mass $b$-baryons will almost certainly decay strongly to
$\Lambda_b\pi$ \cite{rosner}, possibly with large
depolarization \cite{peskin}, some polarization information will
very likely be lost.\par
In a naive spin counting model, the $\Lambda$ and $\Sigma$ baryons have
approximately equal probability of formation \cite{sjostrand},
with higher
mass states being less likely. Thus a 47-94\% polarization of the
$\Lambda_b$ can be reasonably expected  at LEP. A measurement
of polarization will tell us the  degree of fragmentation
into states which retain the $b$-quark polarization relative  to those
which do not.\par
Three
predictions for exclusive $\Lambda_b$
decays have been published \cite{close,korner,schuler}.
However, $b$-baryons are
best observed inclusively at LEP \cite{aleph1}, via an excess of
jets containing a hard lepton and a charge-correlated
$\Lambda_s$, the latter
tagging the cascade
weak decay of the baryon.

Previous suggestions to measure polarization using inclusive semileptonic
$\Lambda_b$ decay \cite{alta} would utilize  only  the electron spectrum and
 were not
sufficiently sensitive to  polarization. Furthermore,
fragmentation uncertainties could compromise the utility of
the previous proposals.

Since there is abundant data (of order a few hundred tagged
$\Lambda_b$'s) to study,
it is  worthwhile  investigating whether
there exists  a variable which is more sensitive to polarization.
In this paper, we show that the variable
 $y=\ev{E_l}/\ev{E_{\bnu}}$
optimizes sensitivity while being remarkably free of theoretical
uncertainties.

Experimental cuts reject events with low-energy leptons and
$\Lambda_s$'s, and therefore reject  most of the already scarce
$b$-hadrons with a fractional
energy less than 0.2 \cite{aleph1}. While this
induces a tiny cut dependence, it helps in two ways. First,
above 0.2, the perturbative depolarizing
effects are very small, and fragmentation and decay
are essentially decoupled, a fact we will use later.
Second,
the relativistic $\beta$ from rest frame to laboratory
is close to unity.
The energy in the laboratory can be expressed as follows:
\[\ev{E}=\ev{\gamma}\ev{E^*}+\ev{\gamma\beta}\ev{p^*}\simeq
\ev{\gamma}( \ev{E^*}+\ev{p^*}),\]
where $p$ is the longitudinal momentum, and all starred quantities
are in the rest frame.
\par
The variable $y$ which we propose is then
\beq y={\ev{E_l} \over \ev{E_{\bnu}}} =
{\ev{E_l^*}+\ev{p_l^*} \over
\ev{E_{\bnu}^*}+\ev{p_{\bnu}^*}}.\eeq
 For  an unpolarized particle,
\beq
y=y_0\equiv{\ev{E_l^*}\over \ev{E_{\bnu}^*}}.
\eeq
\par
Notice that this ratio is dependent only on the rest frame angular
distributions. It is independent of the fragmentation.
We measure sensitivity by the deviation from unity of the
ratio of a fully polarized quantity to the unpolarized value.
We  will   show that with this definition,
 the parameter $y$  is about five times more sensitive
to polarization than the average electron energy alone. The theoretical
uncertainties are at the few percent level.  These
are  the unknown ratio of the charm and beauty quark
masses which we will see gives an effect of this order,
and the deviations from the parton model, which are less than or
of order a few percent.\par
 This is because \cite{chay,shif,wise} the  inclusive differential
distribution from semileptonic hadron decay (appropriately averaged)
is equal to the parton model (free quark) prediction
up to corrections of order $(\Lambda_{QCD}/m_q)^2$ which should
be no more than a few percent.  Notice that by studying the {\it inclusive}
spectrum and focusing
on the lepton system, we do not have the uncertainties
due to  the poorly known Isgur-Wise function which  one has in
the study of  exclusive decays \cite{close,korner,schuler}.

So the rate for
\begin{equation}
b\rightarrow\, c\, l\, \bnu\, ,
\end{equation}
 is proportional to (the $p_x$ here are 4-vectors)
\[ d\Gamma \propto (p_c \cdot p_l)(p_b \cdot p_{\bnu}
-m_b s_b\cdot p_{\bnu})d \Phi\]
where $s_b$ is the spin of the decaying $b$-quark and $d\Phi$ is the
phase space factor. In the unpolarized case, the spin-dependent term
goes to zero.
\par
With the definitions
\begin{eqnarray}
x_l&=&{2 E^*_l \over m_b}\\
x_{\bnu}&=&{2 E^*_{\bnu} \over m_b}\\
\epsilon&=&\left({m_c \over m_b} \right)^2\\
f(\epsilon)&=&1-8\epsilon+8\epsilon^3-
\epsilon^4-12\epsilon^2 {\rm Log}\epsilon
\end{eqnarray}
the inclusive differential decay distribution in the $b$-quark rest frame is
\begin{eqnarray}
{1 \over \Gamma}{d^2\Gamma \over dx_l d \cos\theta_l}&=&{1 \over f(\epsilon)}
{x_l^2(1-\epsilon-x)^2 \over
 (1-x_l)^3}\left((1-x_l)(3-2x_l)+\epsilon(3-x_l)\right. \nonumber \\
&&-\left.\cos\theta_l\left[(1-x_l)(1-2x_l)-\epsilon(1+x_l)\right]\right)
\end{eqnarray}
Here we have defined the angle
of the charged lepton with respect to the direction {\it opposite}
that of the spin in anticipation of the application to a decaying
left handed $b$-quark, where this will be the boost direction.
The inclusive  neutrino  differential decay distribution is predicted to be
\beq
{1 \over \Gamma}{d^2\Gamma \over dx_{\bnu}
d\cos\theta_{\bnu}}={1 \over f(\epsilon)}
(1-x_{\bnu}-\epsilon)x_{\bnu} (1-\cos\theta_{\bnu})
\eeq
The angle was defined with the same convention as that above.

We study only average quantities.
The distributions in the laboratory
are greatly broadened by
the fragmentation function and the $b$-hadron decay.
It can be proven that the
average of each observable is the most sensitive to small differences
independent of the precise knowledge of the fragmentation function
and decay. The average energies and longitudinal momenta
in the fully polarized $b$-quark rest frame
for  100\% polarization  ($P=-1$) are

\begin{eqnarray}
\ev{E_{\bnu}^*}&=&{m_b \over f(\epsilon)}\left({3 \over
 10}-3\epsilon-2\epsilon^2+6\epsilon^3-{3 \epsilon^4 \over 2}
+{\epsilon^5 \over 5}-6\epsilon^2 {\rm Log}\epsilon\right)\nonumber\\
\ev{E_l^*}&=&{m_b
\over f(\epsilon)}\left({7 \over 20}-{15 \epsilon \over 4}
-6\epsilon^2+10\epsilon^3-{3 \epsilon^4 \over 4}+{3 \epsilon^5 \over 20}
-9\epsilon^2 {\rm Log}\epsilon-3\epsilon^3 {\rm Log}\epsilon\right)\nonumber\\
\ev{p_{\bnu}^*}&=&{m_b \over f(\epsilon)}\left(-{1 \over 10}+\epsilon+{2
 \over 3}\epsilon^2-2\epsilon^3+{\epsilon^4 \over 2}-{1\over
 15}\epsilon^5+2\epsilon^2 {\rm Log}\epsilon\right)\nonumber\\
\ev{p_l^*}&=&{m_b
\over f(\epsilon)}\left({1 \over 20}-{3 \over
 4}\epsilon-4\epsilon^2+4\epsilon^3+{3 \over 4}\epsilon^4-{\epsilon^5 \over
 20}-3\epsilon^2{\rm Log} \epsilon-3\epsilon^3{\rm Log}\epsilon\right)\nonumber
\end{eqnarray}
The average energies are independent of polarization.  The average
longitudinal momenta go to zero in the unpolarized case.
This formula is accurate up to corrections of order
$(\Lambda_{QCD}/m_q)^2$
and $(m_l/m_q)^2$.
{}From this we can predict the value of  $y$ as a function of $\epsilon$.
\par
Comparing the four equations above, one can see
that the lepton energy (in the laboratory frame) is about
$7/3$ times less sensitive to polarization than
the neutrino energy. More important,
$y$ is about five times more
sensitive than $\ev{E_l}$ alone. \par
For the purpose of cross-checking the experimental distributions,
we construct also a variable which, unlike $y$, is never singular
event-by-event,
and can be used to check against the experimental
distributions,
\beq
y' = {E_l - E_{\bnu} \over E_l + E_{\bnu}}.
\eeq

\begin{figure}
\let\picnaturalsize=N
\def\picsize{4.0in}
\def\picfilename{lambf2.ps}
\ifx\nopictures Y\else{\ifx\epsfloaded Y\else\input epsf \fi
\let\epsfloaded=Y
\centerline{\ifx\picnaturalsize N\epsfxsize \picsize\fi
 \epsfbox{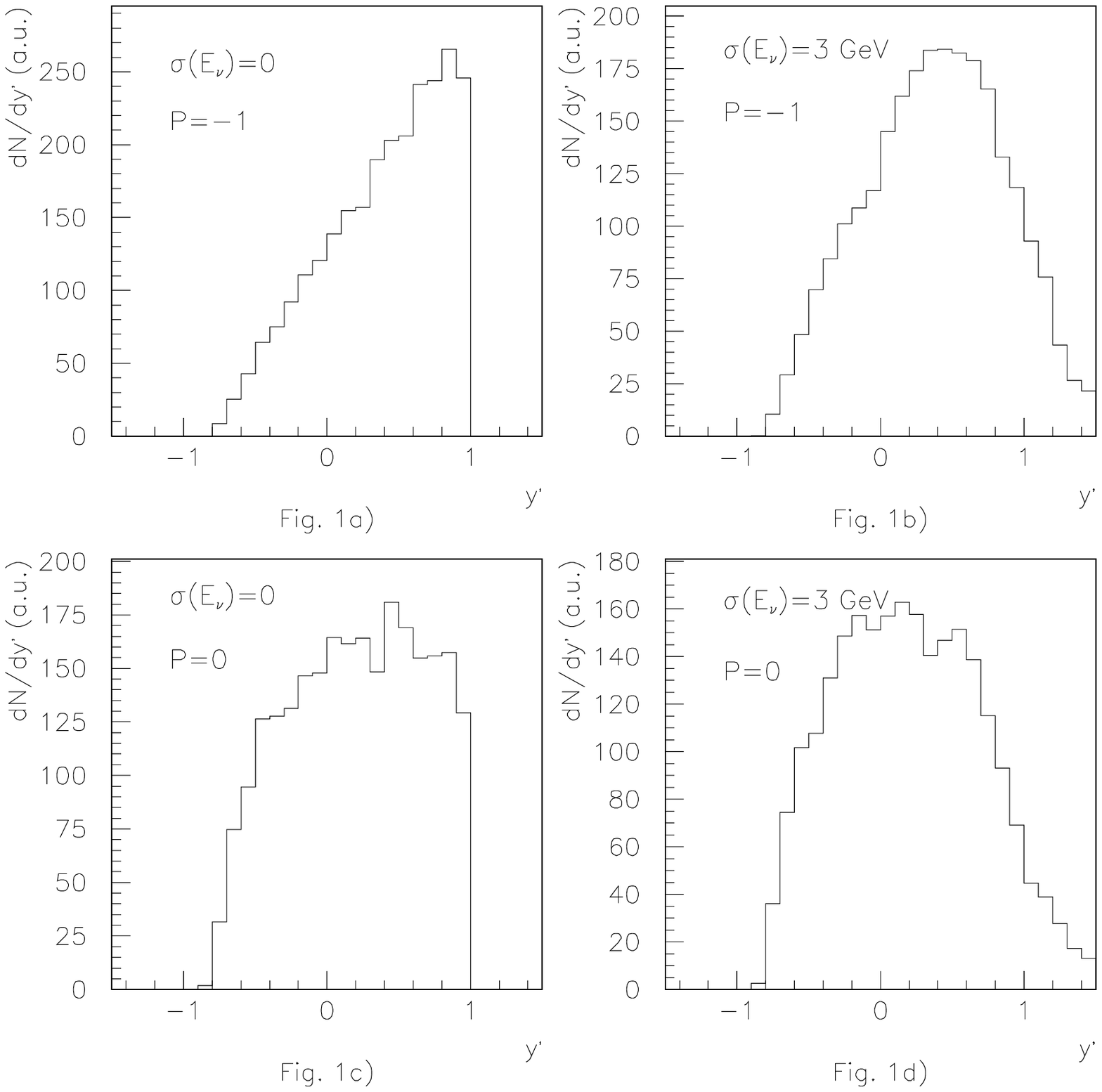}}}\fi
{\figfont  Figure1: The $y'$ distribution, with minimal cuts $E_l>3$GeV,
 $(E_l+E_{\bnu})
>1$GeV, obtained with $10^4$ Monte Carlo events.
a) $P=-1$, perfect neutrino energy
resolution; b) $P=-1$, 3 GeV neutrino energy resolution;
c) $P=0$ perfect neutrino energy resolution; d) $P=0$,
3 GeV neutrino energy resolution.
}
\end{figure}
Fig. 1 shows the distributions for the $y'$ variable. Note that
the  average of the variable
$y'$ is nonzero even  in the unpolarized case. From the figure,
we see that the
realistic cases, Figs. 1b) and 1d), are not
so well distinguished, other than
by their average. The detailed form of the distribution
can be useful however for distinguishing good
from bad events.

\par
Neutrinos can be used at LEP in heavy flavor analysis. If $M_1$ is the
mass of the jet containing the neutrino, $M_2$ the mass of the recoiling
jet, and $E_{beam}$ the beam energy, the neutrino energy
is measured as
\[ E_{\bnu} = {4E^2_{beam}+M_1^2-M_2^2\over 4E_{beam}}-E_{jet}. \]
While this is valid only for neutrinos collinear with the jet,
the formula above exhibits a resolution of
3 GeV \cite{mossad} (entirely due to the jet energy resolution of
the detector),
which is substantially less
than the neutrino energy spectrum (Fig. 2), obtained from
the most recent measurements of the $b$ fragmentation at LEP.
The $y$ resolution is dominated by the neutrino energy resolution.
Comparing 3 GeV with the neutrino energy spectrum
yields an expected $y$ resolution of
($N$ is the number of events)
\[ \sigma_y \sim 0.4/\sqrt{N},\]
where the small coefficient is a non trivial consequence of the
independence of fragmentation and should allow a better than
10\% error with the existing data.

\begin{figure}
\let\picnaturalsize=N
\def\picsize{4.0in}
\def\picfilename{lambf3.ps}
\ifx\nopictures Y\else{\ifx\epsfloaded Y\else\input epsf \fi
\let\epsfloaded=Y
\centerline{\ifx\picnaturalsize N\epsfxsize \picsize\fi
 \epsfbox{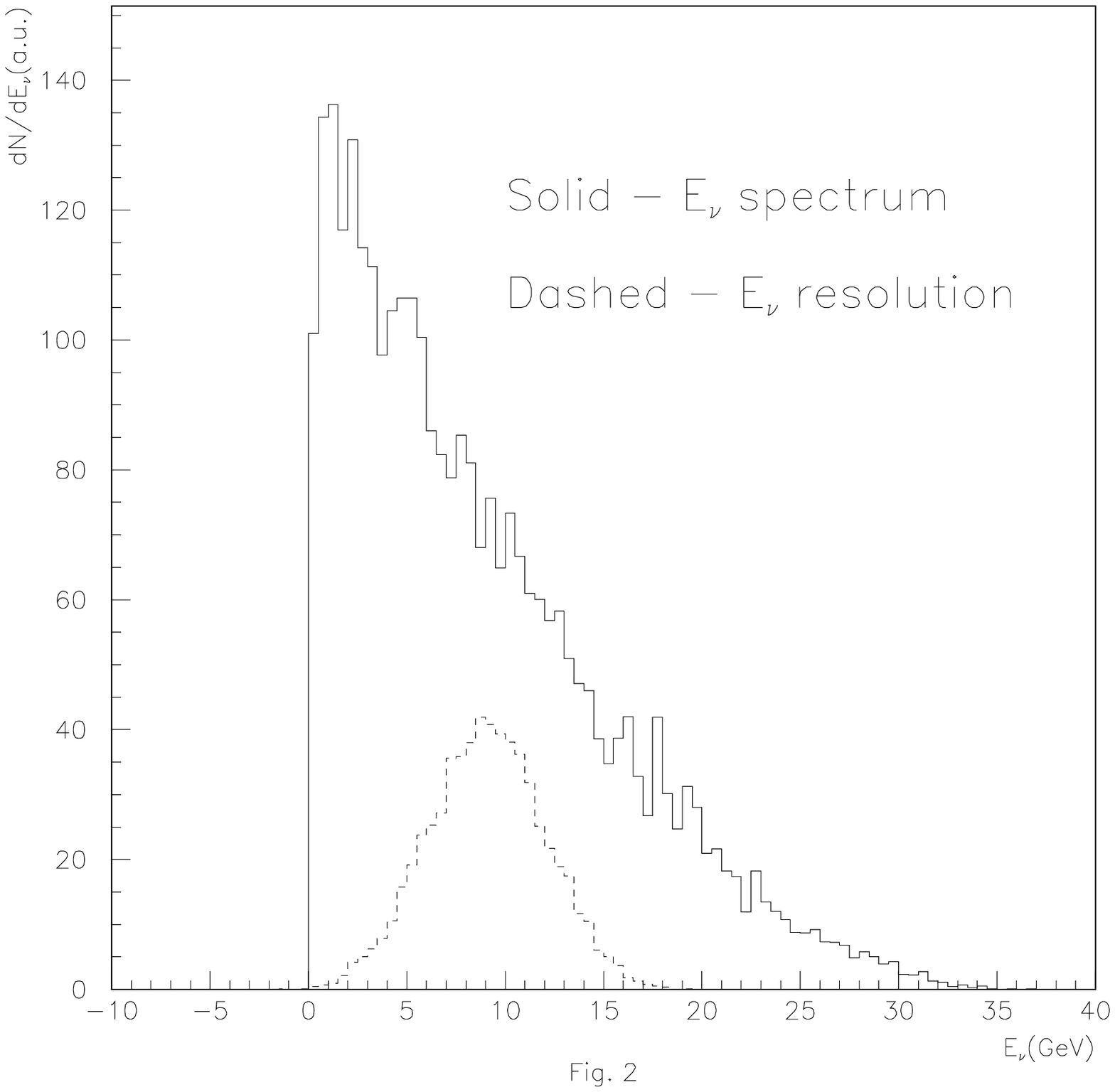}}}\fi
{\figfont  Figure 2: Comparison of the expected neutrino spectrum from
 unpolarized
$b$-hadrons and the
neutrino energy resolution at LEP, as obtained from
$10^4$ Monte Carlo events.
Solid: neutrino energy
spectrum; dashed: expected neutrino energy resolution (3 GeV)
for a 9 GeV
neutrino.
}
\end{figure}

\par
Practically, it will be easier to take the difference or ratio between
a $b$-meson sample and a $b$-baryon tagged sample, as this
eliminates a  plethora of systematics.
The ratio and difference of the baryon and meson samples,
\[ R_y=y_{baryon}/ y_{meson}, \;\;\; D_y = y_{baryon}-y_{meson}
\]
are well calibrated with respect to polarization
\begin{equation}
R_y  =  1-K_R P, \; \; \; \;
D_y  =  - K_{D}P
\end{equation}
with $K_R = 0.66\pm 0.02$ and $K_{D} = 0.75\pm 0.03$,
for $\epsilon$ ranging between 0.06 and 0.14, and before cuts.
The difference from 1 and 0 respectively is
a direct measure of polarization.
Neither QCD corrections, nor $m_c$ mass
uncertainties (which enter in the error purely due to kinematics),
nor the kinematical approximations used above contribute
more than 0.02 to the calibration error.
\par
As a cross check, we consider
the average neutrino and lepton energies
themselves rather than the ratio or difference. Here
one can consider qualitative questions such as whether
the spectrum of the leptons from baryons or mesons is harder.
To address more quantitative questions one needs an
accurate measurement of the baryon and meson fragmentation spectrum,
since this affects the overall energy scale. Recall that the
mean value of the hadron
energy, $\ev{z}=\ev{E}/E_{beam}$,
factorizes into a perturbative and nonperturbative
contribution, that is,
\beq
\ev{z}=\ev{z}_{pert}\ev{z}_{nonpert}
\eeq
 Of course the
exact factorization is renormalization
scale dependent. If one takes
the renormalization scale to be of order
the heavy quark mass, $\ev{z}_{nonpert}$ is
intrinsically nonperturbative, but  for sufficiently
large $m_q$ can
be expanded as
\beq
\ev{z}_{nonpert}= 1-a({\Lambda_{QCD}\over m_q})+
O(({\Lambda_{QCD}\over m_q})^2),
\eeq
where $a>0$ depends on the hadron type \cite{jr,nason}.
The first factor, $\ev{z}_{pert}$ is
independent of the hadron type and is
determined by Altarelli-Parisi evolution,
but $\ev{z}_{nonpert}$ depends on hadron type.  In principle,
$\ev{z}_{nonpert}$ for the different hadron types
can differ by an amount of order $\Lambda_{QCD}/m_q$ which could
disguise polarization effects such as a difference between the lepton
spectra in the two samples.
\par
In principle,
one can use perturbative QCD and the heavy quark expansion
to predict the $b$-quark fragmentation parameters,
but with fairly large QCD uncertainties \cite{jr,nason,mn}. However,
one can use directly the measured fragmentation functions
of $b$ hadrons from LEP. This is justified because the mean $\ev{z}$
for the meson and baryon can be shown to be very nearly the
same, based on the ARGUS charmed particle
results \cite{argus}, just below $B$ threshold,
which are in agreement with preliminary results from CLEO \cite{cleo}.
\par

Within a 2.5\% error, the ARGUS results
indicate that the $\ev{z}$ of the $D$, $D^*$ and $\Lambda_c$
are the same. The measured
value is about $0.65$.
\par
Therefore,
\beq
(\ev{z}_{baryon}-\ev{z}_{meson})< a_b-a_m {\Lambda_{QCD} \over m_b}\le
{0.025 \over 0.65}{m_c
 \over m_b}\approx 1.25 \cdot 10^{-2}
\eeq
Here we have taken the difference between $c$-meson and $c$-baryon
mean $\ev{z}$ to be of order 2.5\% and have taken the maximum perturbative
contribution to be the value which is measured (since $a$ is positive)
which is approximately $0.65$.
Finally, we have scaled the nonperturbative correction by the ratio
of quark masses, since we have measured $c$ quark fragmentation
but need to predict that for the $b$-quark. We conclude
that a $2.5\%$ agreement in the $D$ system must translate conservatively
into $\ev{z_B}\approx\ev{z_{\Lambda_b}}$ at  the percent level
which is well within the theoretical uncertainty.
\par
With this result in hand,
one can predict  that the left handedness of the $\Lambda_b$ will manifest
itself in a slightly higher $\ev{E_l}$ and a substantially lower
$\ev{E_{\bnu}}$ compared to the meson sample, which is a useful
cross check.
\par
In conclusion, this paper has addressed the problem of observing
polarization transfer in hadronic $Z$ decays. It was found that
sensitive, model-independent variables can be extracted from the
lepton-neutrino system.
The fragmentation
problem was solved in two different ways. Our proposed
variable $y$ is almost free of theoretical error and increases the
sensitivity to polarization by at least a factor of five compared
to previous proposals.
\par
Once the $b$-polarization is measured, new
 information will be obtained about the relative fragmentation
into the $\Sigma_b$ and $\Lambda_b$ baryons.
Later, it should be possible
to tackle $c$ polarization, and both measurements together should
provide information on the spin structure of QCD in the non-perturbative
regime. \par
We thank Bob Cahn, Cristi Diaconu,
Mike Luke, Paolo Nason, Marcello Quattromini,
Nuria Rius and Mossadek Talby
for helpful discussions. L. R. thanks the CERN theory group for
their hospitality when this work was initiated.

\vspace{0.5in}
\Large
\end{document}